\documentclass[cits]{PoS}
\usepackage{lineno}
\title{J/$\psi$ and $\psi$(2S) production in p-Pb collisions with ALICE at the LHC}

\ShortTitle{J/$\psi$ and $\psi$(2S) production in p-Pb collisions with ALICE at the LHC}

\author{\speaker{Marco LEONCINO} for the ALICE Collaboration%
        \\
        INFN and University, Torino\\
        E-mail: \email{leoncino@to.infn.it}}

\abstract{
The ALICE collaboration has studied the inclusive J/$\psi$ and $\psi(\mathrm{2S})$ production in p-Pb collisions at $\sqrt{s_{NN}}$ = 5.02 TeV at the CERN LHC.
The J/$\psi$ measurement is performed in the $\mu^{+}\mu^{-}$ ( - 4.46 < $y_{cms}$ < - 2.96  and  2.03 < $y_{cms}$ < 3.53 ) and in the $e^{+}e^{-}$ ( - 1.37 < $y_{cms}$ < 0.46 ) decay channels, 
down to zero transverse momentum. The results are in fair agreement with theoretical predictions based on nuclear shadowing, as well as with models including, in addition, a contribution from partonic 
energy loss. Finally, the $\psi(\mathrm{2S})$ measurement in the $\mu^{+}\mu^{-}$ decay channel has been performed. In particular, a significantly smaller $\psi(\mathrm{2S})$ nuclear 
modification factor, with respect to the J$/\psi$ one, has been observed.
}

\FullConference{52 International Winter Meeting on Nuclear Physics\\
                 27 - 31 January 2014\\
                 Bormio, Italy}

\begin{document}


\section{Introduction}
The suppression of charmonia, bound states of \textit{c} and \textit{$\bar{c}$} quarks, was proposed by T. Matsui and H. Satz \cite{Matsui_Satz} as a clean signature of 
Quark-Gluon-Plasma (QGP) formation in heavy-ion collisions. The available data from SPS \cite{SPS1,SPS2} and RHIC \cite{RHIC1,RHIC2} are showing ``anomalous-suppression'',
which could be indicating the charmonia suppression. In this situation, ALICE results \cite{ALICE1}, which show smaller J/$\psi$ suppression compared to lower energy experiments,
indicate clearly the importance of a process for creating J/$\psi$ in case of heavy ion collisions at LHC energy.
Partonic transport models \cite{Partonic1,Partonic2} as well as statistical generation models \cite{Statistical_Regeneration}, 
explain this effect as caused by a regeneration of J/$\psi$ along the collision history and/or hadronization, favoured by the large \textit{c}\textit{$\bar{c}$} multiplicity typical at the LHC energies.
In addition to the color screening mechanism, it became soon clear that other effects may contribute to the charmonium suppression. In particular, nuclear effects (shadowing \cite{Shadowing1,Shadowing2,
Shadowing3,Shadowing4,Shadowing5} and initial state parton energy loss \cite{Eloss1,Eloss2,Eloss3}) are expected to play a role, contributing to the suppression of charmonium states. 
For this reason it is clear that is very important to study the production of charmonium in pA collisions, since it gives access to nuclear effects and it consequently allows to disentangle the suppression 
contribution related to them (cold nuclear matter effects) from that associated to the formation of a QGP (hot nuclear matter effects). 

\section{ALICE detector and data taking conditions}
ALICE (A Large Ion Collider Experiment) is the dedicated heavy-ion detector at the LHC. The detectors of the central region ($\left|\eta\right|< 0.9$) are embedded in a large solenoidal magnet 
with a maximum field of 0.5 T. The main tracking devices are the Inner Tracking System (ITS), made of six layers of silicon detectors around the beam pipe, and the Time Projection Chamber (TPC), a large
cylindrical gas detector. The TPC is also used to identify particles via the measurement of the energy loss $dE/dx$. The Muon Spectrometer, which covers the forward region ($-4<\eta<-2.5$), is composed of a set
of absorbers, a 3 $\mathrm{T\cdot m}$ dipole magnet, ten planes of tracking chambers and four planes of trigger chambers. Two VZERO hodoscopes ($2.8<\eta<5.1$ and $-3.7<\eta<-1.7$) are used for triggering
and the Zero Degree Calorimeters (ZDC), placed 112.5 m from the interaction point, are used to reject satellite collisions. More details about the experimental setup can be found in \cite{ALICE_detector}.
In this analysis the J/$\psi$ resonance is detected both in the dielectron decay channel (using the central barrel detectors) and in the dimuon decay channel 
(using the forward muon spectrometer), while the $\psi(\mathrm{2S})$ is detected only in the dimuon decay channel.
Due to the energy asymmetry of the LHC beams in p-Pb collisions, ($E_\mathrm{{p}}$ = 4 TeV, $E_\mathrm{{Pb}}$ = 1.58 $\cdot A_\mathrm{{Pb}}$ TeV, where $A_\mathrm{{Pb}} = 208$) the nucleon-nucleon center-of-mass system 
doesn't coincide with the laboratory system, but is shifted by $\Delta y = 0.465$ in the direction of the proton beam. This results in asymmetric rapidity coverages with respect to y=0: -4.46 < $y_{cms}$ < -2.96 at 
backward rapidity, -1.37 < $y_{cms}$ < 0.46 at midrapidity and 2.03 < $y_{cms}$ < 3.53 at forward rapidity. In the case of the $e^{+}e^{-}$ decay channel the presented results are based on a Minimum Bias trigger 
($L_\mathrm{{int}}$ = 52 $\mu b^{-1}$) and on a dimuon trigger in the case of the $\mu^{+}\mu^{-}$ decay channel with ($L_\mathrm{{int}}$ = 5.0 (5.8) $\mu b^{-1}$) at forward (backward) rapidity respectively.

\section{Data analysis}
The signal extraction is performed by analyzing the invariant mass spectra, both in the $e^{+}e^{-}$ and in the $\mu^{+}\mu^{-}$ channel. In the $e^{+}e^{-}$ channel the signal extraction is performed after 
subtracting the combinatorial background, obtained from the like-sign spectrum, normalized to match the integral of the opposite-sign dielectron in the invariant mass region 3.2-4.9 GeV/$c^{2}$.
The signal events are then obtained by bin counting in the invariant mass region 2.92-3.16 GeV/$c^{2}$.
In the dimuon decay channel, the J/$\psi$ and $\psi(\mathrm{2S})$ signals are extracted with a fit of the opposite sign mass spectrum, using a Crystal Ball function \cite{CB2_function} or a pseudo-gaussian 
functions \cite{NA60_function} for the resonances and a variable-width-gaussian or the product of a $4^{th}$ order-polynomial and an exponential for the background. 
Figures 1 and 2 show the invariant mass spectra and the results of the signal extraction, in the dielectron and in the dimuon decay channel respectively.

\begin{figure}[htbp]
\begin{center}
\includegraphics[width=7.8cm]{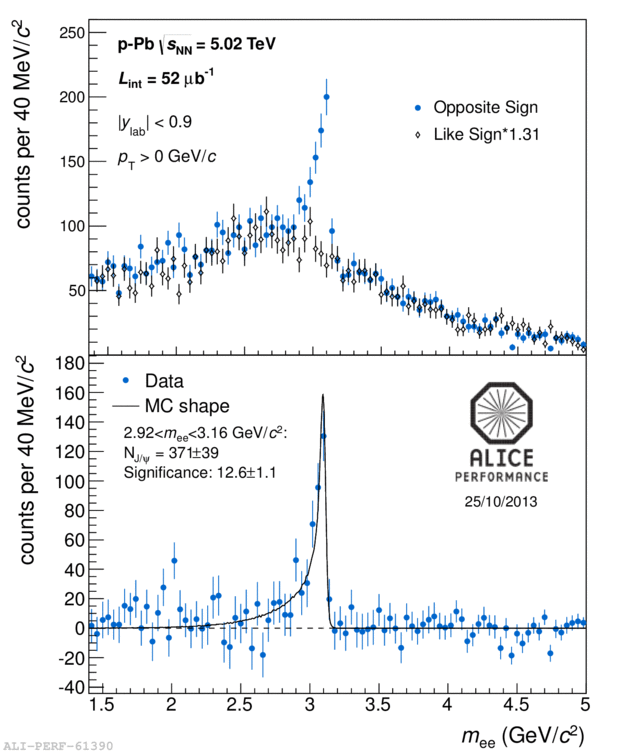}
\caption{Opposite-sign dielectron invariant mass spectrum at midrapidity. }
\label{}
\end{center}
\end{figure}

\begin{figure}[htbp]
\begin{center}
\includegraphics[width=7.1cm]{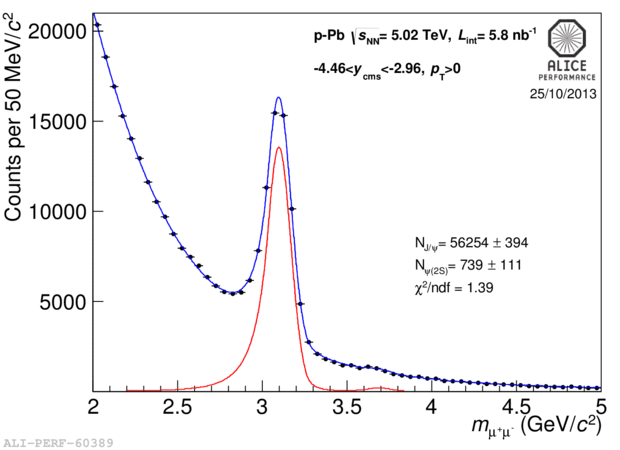}
\qquad
\includegraphics[width=7.1cm]{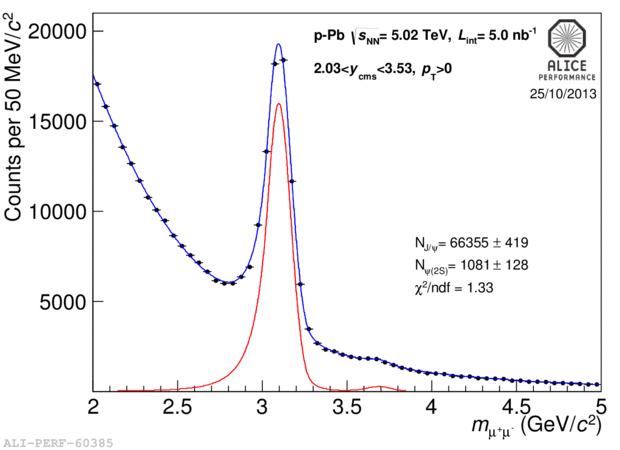}
\caption{Opposite-sign $\mu^{+}\mu^{-}$ invariant mass spectra at backward (left) and forward (right) rapidity. The fits shown in these plots are obtained with an extended Crystal Ball function 
for the signal (shown as a red line) and a variable width gaussian for the background.}
\label{}
\end{center}
\end{figure}

\section{Results}
\subsection{J/$\psi$}

The nuclear effects on J/$\psi$ production in p-Pb collisions are quantified using the nuclear modification factor $R_{\mathrm{pPb}}$, which, in the J/$\psi$ $\rightarrow$ $\mu^{+}\mu^{-}$ decay channel, is defined by:

\begin{displaymath}
R_{\mathrm{pPb}}=\frac{N_{\mathrm{J/\psi}\rightarrow \mu\mu}^{\mathrm{cor}}}{\left \langle T_\mathrm{{pPb}} \right \rangle \cdot \mathrm{N_{MB}}\cdot \mathrm{B.R.}
(\mathrm{J/\psi}\rightarrow\mu\mu)\cdot\mathrm{\sigma_{pp}^{J/\psi}}}
\end{displaymath}

where $\mathrm{\sigma_{pp}^{J/\psi}}$ is the production cross section in pp collisions in the same kinematical range at the same energy and $\left \langle T_\mathrm{{pPb}} \right \rangle$ is the nuclear
thickness function estimated through the Glauber model \cite{T_pPb}. Since pp data at $\sqrt{s}$ = 5.02 TeV are not available, the reference  $\mathrm{\sigma_{pp}^{J/\psi}}$ is obtained with an interpolation 
procedure. In case of the results in the dimuon decay channel, the pp-reference for the nuclear modification factor relies on an interpolation of the ALICE measurements at
$\sqrt{s}$ = 2.76 TeV and 7 TeV in the muon spectrometer and is performed bin-by-bin in y or $p_\mathrm{{T}}$. For the rapidity bins not covered due to the rapidity shift, an extrapolation is used. More details
about the explored method as a function of rapidity can be found in \cite{CS_interpolation}. For the dielectron decay channel, the reference cross section was evaluated based on the interpolation of $ \mathrm{BR} \times d\sigma /dy$
measurements in pp ({$\mathrm{p\bar{p}}$}) collisions at central rapidity published by the PHENIX Collaboration at $\sqrt{s}$ = 200 GeV \cite{interpol_ele1}, by the CDF Collaboration at $\sqrt{s}$ = 1960 GeV \cite{interpol_ele2} and 
the ALICE Collaboration at $\sqrt{s}$ = 2.76 TeV \cite{interpol_ele3} and $\sqrt{s}$ = 7 TeV \cite{interpol_ele4}. The J/$\psi$ $R_{\mathrm{pPb}}$, integrated over $p_\mathrm{{T}}$, is shown in Figure 3.

\begin{figure}[h!]
\begin{center}
\includegraphics[width=9.5cm]{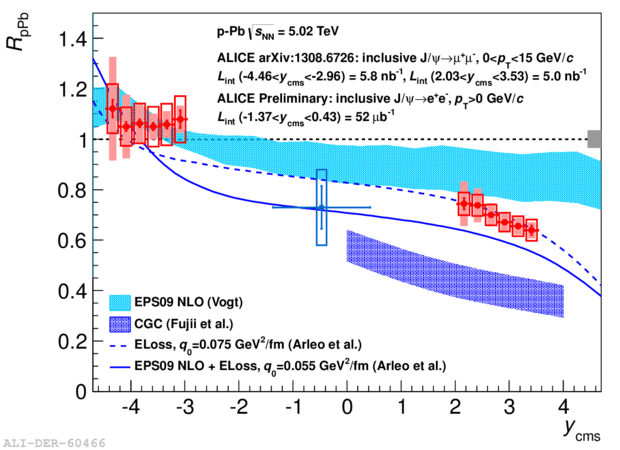}
\caption{Inclusive J/$\psi$ nuclear modification factor as a function of rapidity. Open boxes are uncorrelated systematic uncertainties, filled area indicate partially correlated uncertainties and 
the gray box the global uncertainty due to the $T_\mathrm{{pPb}}$. Theoretical models are also shown.}
\label{}
\end{center}
\end{figure}

At mid and forward rapidity, the inclusive J/$\psi$ production is suppressed with respect to that one in binary-scaled pp collisions, whereas it is unchanged at backward rapidity. ALICE results are 
compared to theoretical models; ones with a EPS09 PDF \cite{EPS09_1,EPS09_2}, one with coherent parton energy loss \cite{ElossAlice} and one with the Color Glass Condensate (CGC) initial state \cite{CGCAlice}. 
Models containing shadowing and/or energy loss are in agreement with ALICE data (within uncertainties), while the CGC model overestimates the suppression. In Figure 4 the J/$\psi$ $R_{\mathrm{pPb}}$ is shown 
as a function of $p_\mathrm{{T}}$, in the three rapidity intervals accessible by ALICE.

\begin{figure}[h!]
\begin{center}
\includegraphics[width=6.8cm]{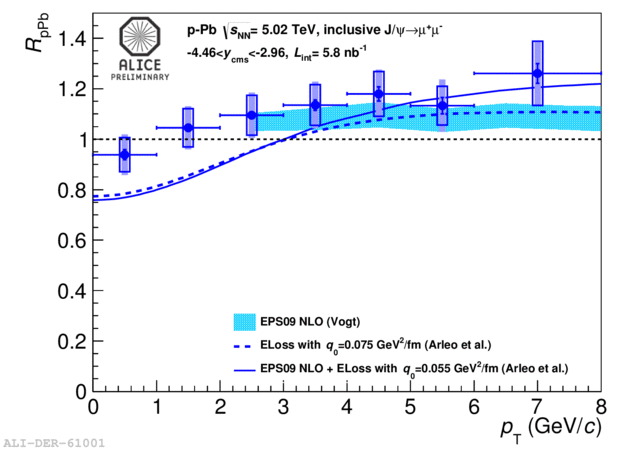}
\qquad
\includegraphics[width=6.8cm]{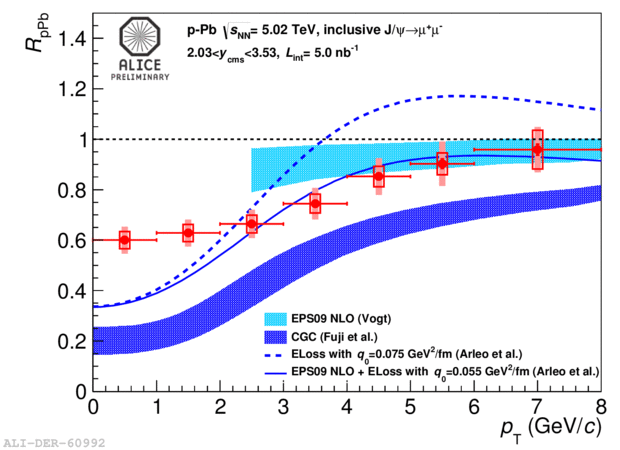}
\qquad
\includegraphics[width=7.2cm]{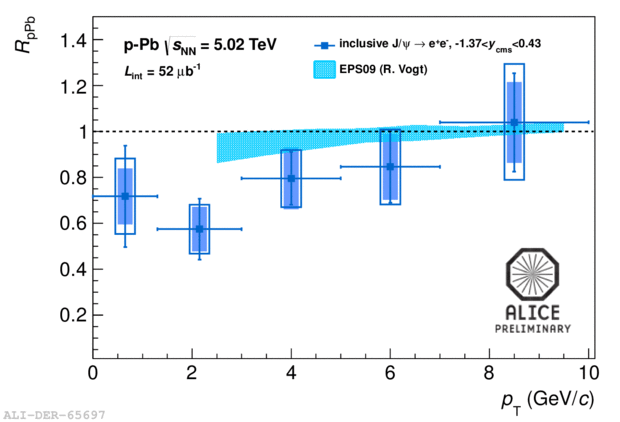}
\qquad
\caption{J/$\psi$ $R_{\mathrm{pPb}}$ as a function of $p_\mathrm{{T}}$ at backward (top-left), forward (top-right) and mid (bottom-center) rapidity. Open boxes are uncorrelated systematic uncertainties, filled area indicate 
partially correlated uncertainties. The results of theory calculations are also shown.}
\label{}
\end{center}
\end{figure}

The $p_\mathrm{{T}}$ dependence of the J/$\psi$ nuclear modification factor is more pronounced at forward than at backward rapidity. Comparison with the model
calculations seems to show that any of the models could not reproduce the experimental trend over the whole $p_\mathrm{{T}}$ comfortably. Since the Bjorken x-values in the Pb nucleus in p-Pb collisions 
at $\sqrt{s_{NN}}$ = 5.02 TeV are similar to the ones in Pb-Pb collisions at $\sqrt{s_{NN}}$ = 2.76 TeV, assuming a factorization of shadowing effects, it is possible to estimate the expected size of cold nuclear matter
effects in Pb-Pb. This is done by comparing $R_{\mathrm{pPb}}$ in p-Pb collisions with the Pb-Pb data.
In particular, under the previous hypothesis, the J/$\psi$ $R_\mathrm{{pPb,backward}} \cdot R_\mathrm{{pPb,forward}} (R_\mathrm{{pPb, midrapidity}}^{2})$ is compared to the 
$R_\mathrm{{PbPb, forward}}(R_\mathrm{{PbPb, midrapidity}})$. These results are shown in Figure 5.

\begin{figure}[h!]
\begin{center}
\includegraphics[width=6.8cm]{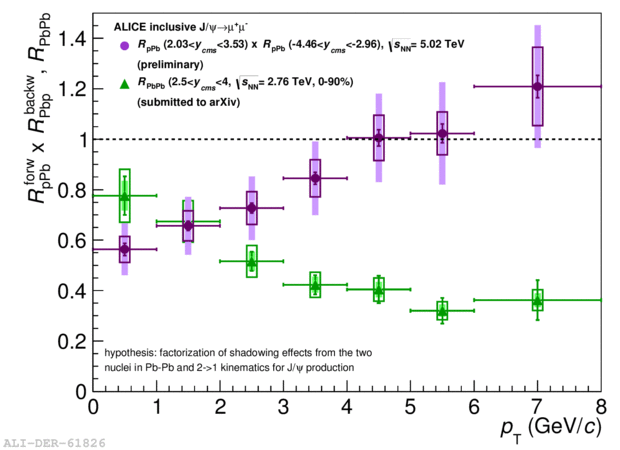}
\qquad
\includegraphics[width=7.2cm]{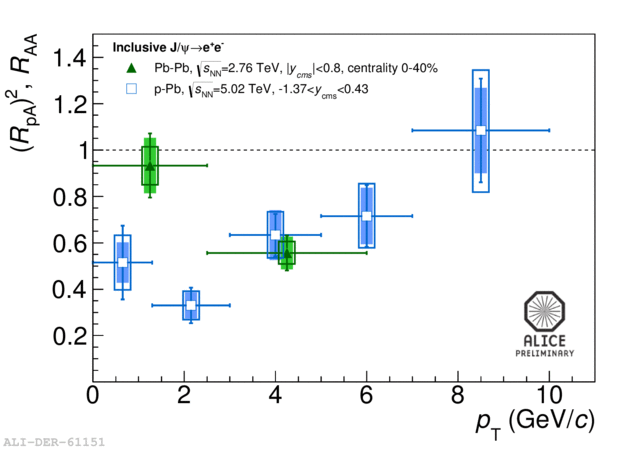}
\caption{In the left plot the J/$\psi$ $R_\mathrm{{pPb,backward}}\cdot R_{\mathrm{pPb,forward}}$ is compared to the $R_\mathrm{{pPbPb,forward}}$. In the right plot the J/$\psi$ $R_\mathrm{{pPb,midrapidity}}^{2}$ is compared
with $R_\mathrm{{PbPb,midrapidity}}$. Systematic uncertainties are represented in the same way as Figure 4.}
\label{}
\end{center}
\end{figure}

\vspace{0.5cm}

In the muon decay channel it is possible to introduce the variable $S_{J/\psi}$ which is given by:

\begin{displaymath}
S_{J/\psi}=\frac{R_\mathrm{{PbPb}}}{(R_\mathrm{{pPb}}^\mathrm{{forward}}\cdot R_\mathrm{{pPb}}^\mathrm{{backward}})}
\end{displaymath}

The $S_\mathrm{{J/\psi}}$ variable, for the inclusive ${J/\psi}\rightarrow \mu^{+}\mu^{-}$ is shown, as a function of $p_\mathrm{{T}}$, in Figure 6.

\begin{figure}[h!]
\begin{center}
\includegraphics[width=8cm]{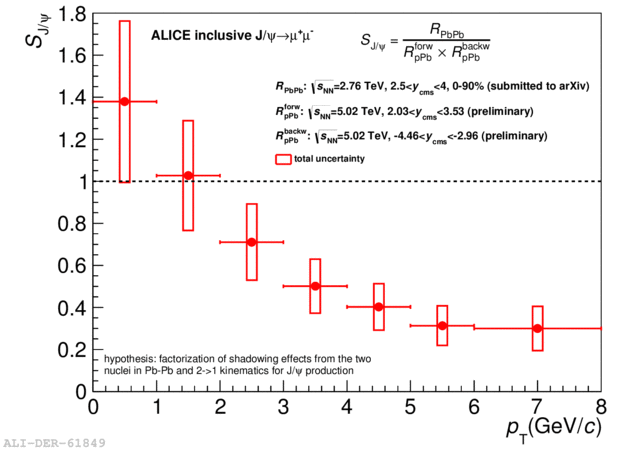}
\caption{The $S_{J/\psi}$ variable for the inclusive ${J/\psi}\rightarrow \mu^{+}\mu^{-}$ }
\label{}
\end{center}
\end{figure}

The comparison between the suppression pattern in p-Pb and Pb-Pb collisions presented in Fig.6, shows that, at low transverse momentum, there could be a contribution from regeneration (this results is in 
agreement with partonic transport models \cite{Partonic1}) while, at higher transverse momenta, the suppression contribution starts to be dominant.

\subsection{$\psi(\mathrm{2S})$}
The $\psi(\mathrm{2S})$ analysis at backward and forward rapidity has been performed analogously to the J/$\psi$ one. In Figure 7 the double ratio $\mathrm{\left[\psi(2S))/J/\psi\right]_{pPb}/\left[\psi(2S))/J/\psi\right]_{pp}}$
is shown as a function of rapidity and it is compared with the PHENIX results in d-Au collisions at $\sqrt{s_{NN}}$ = 0.2 TeV. 
ALICE results show a similar trend compared with PHENIX data at midrapidity \cite{PHENIX} showing a decrease of the $\psi(\mathrm{2S})/J/\psi$ ratio in p-Pb collisions with respect to the one in pp collisions. 

\begin{figure}[h!]
\begin{center}
\includegraphics[width=8.8cm]{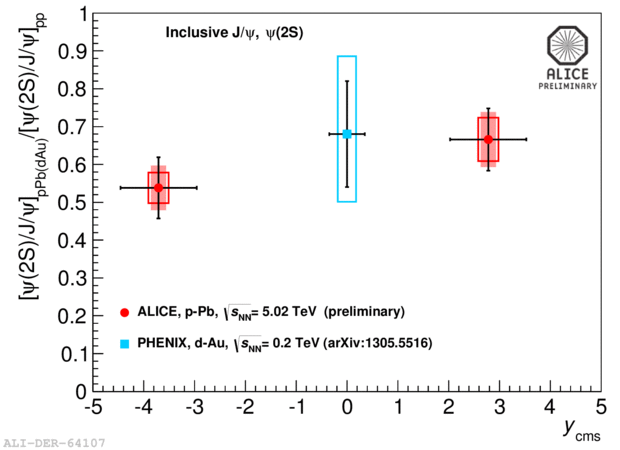}
\caption{The double ratio $\mathrm{\left[\psi(2S))/J/\psi\right]_{pPb}/\left[\psi(2S))/J/\psi\right]_{pp}}$. PHENIX data at $\sqrt{s_{NN}}$ = 0.2 TeV at midrapidity is also shown. The shaded area represents partially
correlated systematics while the boxes are correlated systematics.}
\label{}
\end{center}
\end{figure}

\begin{figure}[h!]
\begin{center}
\includegraphics[width=8.8cm]{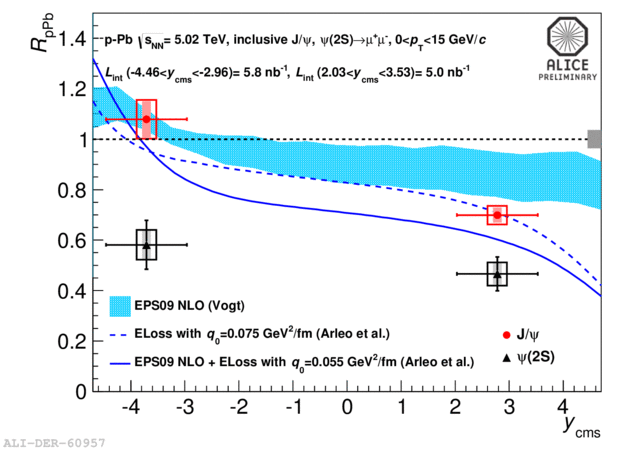}
\caption{The $R_{\mathrm{pPb}}^{\psi(2S)}$ compared to $R_{\mathrm{pPb}}^{J/\psi}$. Open boxes are uncorrelated systematic uncertainties, filled area indicate partially correlated uncertainties and 
the gray box the global uncertainty. Theoretical models for the J/${\psi}$ \cite{EPS09_1,EPS09_2,ElossAlice}, which are also valid for $\psi(\mathrm{2S})$ are shown.}
\label{}
\end{center}
\end{figure}

The $\psi(\mathrm{2S})$ nuclear modification factor $R_{\mathrm{pPb}}$, can be obtained in the following way:

\begin{displaymath}
R_{\mathrm{pPb}}^{\psi(2S)}=R_{\mathrm{pPb}}^{J/\psi}\cdot \frac{\sigma_{\mathrm{pPb}}^{\psi(2S)}}{\sigma_{\mathrm{pPb}}^{J/\psi}}\cdot \frac{\sigma_{\mathrm{pp}}^{J/\psi}}{\sigma_{\mathrm{pp}}^{\psi(2S)}}
\end{displaymath}

In Figure 8 the $R_{\mathrm{pPb}}^{\psi(2S)}$ is presented as a function of rapidity and it is compared to $R_{\mathrm{pPb}}^{J/\psi}$.

The $\psi(\mathrm{2S})$ is more suppressed with respect to the J/${\psi}$. These results are compared to theoretical calculations used for the J/${\psi}$ \cite{EPS09_1,EPS09_2,ElossAlice}, 
which provide almost identical for the $\psi(\mathrm{2S})$ (shadowing effects are supposed to be identical for J/${\psi}$ and $\psi(\mathrm{2S})$ because of the similar kinematic distributions of gluons and, for what concern coherent
energy loss, no sensitivity to the final quantum numbers of the charmonium state is expected). The available theoretical predictions do not describe the $\psi(\mathrm{2S})$ suppression: other
mechanisms are required to explain it.

\section{Conclusions}

In summary, the ALICE collaboration has studied the J/${\psi}$ and $\psi(\mathrm{2S})$ production in p-Pb collisions at $\sqrt{s_{NN}}$ = 5.02 TeV. The J/${\psi}$ nuclear modification factor has been
shown as a function of rapidity and transverse momentum. Results are in fair agreement with theoretical models based on shadowing and/or energy loss, while the Color Glass Condensate approach 
overestimates the suppression. The $\psi(\mathrm{2S})$, studied as a function of rapidity, is more suppressed with respect to the J/$\psi$ and the available calculation do not describe the observed 
nuclear modification factor.


\begin{thebibliography}{99}
\bibitem{Matsui_Satz} T. Matsui and H. Satz., \emph{J/{$\psi$} Suppression by Quark-Gluon Plasma Formation}, \emph{Phys. Lett.} {\bf B 178} (1986) 416 [{\tt INSPIRE}].
\bibitem{SPS1} B. Alessandro et al. (NA50 Collaboration), \emph{A new measurement of J/{$\psi$} suppression in Pb-Pb collisions at 158 GeV per nucleon}, \emph{Eur. Phys. J.}, {\bf C 39} (2005) 335 {\tt [hep-ex/0412036] [INSPIRE]}.
\bibitem{SPS2} R. Arnaldi et al. (NA50 Collaboration), \emph{J/{$\psi$} production in indium-indium collisions at 158-GeV/nucleon}, \emph{Phys. Rev. Lett.}, {\bf 96} (2007) 132302 [{\tt INSPIRE}].
\bibitem{RHIC1} A. Adare et al. (PHENIX Collaboration), \emph{J/{$\psi$} suppression at forward rapidity in Au+Au collisions at $\sqrt{s_{NN}}$ = 200 GeV}, \emph{Phys. Rev.}, {\bf C 84} (2011) 054912 {\tt [arXiv:1103.6269] [INSPIRE]}.
\bibitem{RHIC2} L. Adamczyk et al. (STAR Collaboration), \emph{J/{$\psi$} production at high transverse momenta in p+p and Au+Au collisions at $\sqrt{s_{NN}}$ = 200 GeV}, \emph{Phys. Lett.}, {\bf B 722} (2013) [{\tt arXiv:1208.2736}].
\bibitem{ALICE1} B. Abelev et al. (ALICE Collaboration), \emph{J/{$\psi$} Suppression at Forward Rapidity in Pb-Pb Collisions at $\sqrt{s_{NN}} = \mathrm{2.76 TeV}$ }, \emph{Phys. Rev. Lett.}, {\bf109} (2012) 072301. 
\bibitem{Partonic1} X. Zhao and R. Rapp, \emph{Medium Modifications and Production of Charmonia at LHC}, \emph{Nucl. Phys.} {\bf A 859} (2011) 114 [{\tt arXiv:1102.2194}].
\bibitem{Partonic2} Y. Liu, Z. Qu, N. Xu and P. Zhuang, \emph{J/{$\psi$} Transverse Momentum Distribution in High Energy Nuclear Collisions}, \emph{Phys. Lett.}, {\bf B 678} (2009) 72 [{\tt arXiv:0901.2757}].
\bibitem{Statistical_Regeneration} A. Andronic, P. Braun-Munzinger,K. Redlich and J. Stachel, \emph{The thermal model on the verge of the ultimate test: particle production in Pb-Pb collisions at the LHC}, \emph{J. Phys.} {\bf G 38}, (2011) 124081.
\bibitem{Shadowing1} J. Aubert et al. (European Muon collaboration), \emph{The ratio of the nucleon structure functions $F2_{n}$ for iron and deuterium}, \emph{Phys. Lett.} {\bf B 123} (1983) 275 [{\tt INSPIRE}].
\bibitem{Shadowing2} K. Eskola, H. Paukkunen and C. Salgado, \emph{EPS09: A New Generation of NLO and LO Nuclear Parton Distribution Functions}, \emph{JHEP} {\bf 04} (2009) 065 {\tt [arXiv:0902.4154] [INSPIRE]}.
\bibitem{Shadowing3} D. de Florian, R. Sassot, P. Zurita and M. Stratmann, \emph{Global Analysis of Nuclear Parton Distributions}, \emph{Phys. Rev.} {\bf D 85} (2012) 074028 {\tt[arXiv:1112.6324] [INSPIRE]}.
\bibitem{Shadowing4} D. de Florian and R. Sassot, \emph{Nuclear parton distributions at next-to-leading order}, \emph{Phys. Rev.} {\bf D 69} (2004) 074028 {\tt [hep-ph/0311227] [INSPIRE]}.
\bibitem{Shadowing5} M. Hirai, S. Kumano and T.-H. Nagai, \emph{Determination of nuclear parton distribution functions and their uncertainties in next-to-leading order}, \emph{Phys. Rev.} {\bf C 76} (2007) 065207 {\tt[arXiv:0709.3038] [INSPIRE]}.
\bibitem{Eloss1} S. Gavin and J. Milana, \emph{Energy loss at large $x_F$ in nuclear collisions}, \emph{Phys. Rev. Lett} {\bf 68} (1992) 1834 [{\tt INSPIRE}].
\bibitem{Eloss2} S.J. Brodsky and P. Hoyer, \emph{A Bound on the energy loss of partons in nuclei}, \emph{Phys. Lett.} {\bf B 298} (1993) {\tt[hep-ph/9210262] [INSPIRE]}.
\bibitem{Eloss3} F. Arleo and S. Peigne, \emph{J/$\psi$ suppression in p-A collisions from parton energy loss in cold QCD matter}, \emph{Phys. Rev. Lett.} {\bf 109} (2012) 122301 {\tt[arXiv:1204.4609] [INSPIRE]}.
\bibitem{ALICE_detector} K. Aamodt et al. (ALICE Collaboration), \emph{JINST}, {\bf 3} (2008) S08002.
\bibitem{CB2_function} J. Gaiser, \emph{Charmonium spectroscopy from radiative dacays of the J/$\psi$ and $\psi$'}, Ph.D. Thesis, SLAC Stanford (1982), SLAC-255 (82.REC.JUN.83) {\tt[http://www.slac.stanford.edu/cgi-wrap/getdoc/slac-r-255.pdf]}.
\bibitem{NA60_function} R. Shahoyan, \emph{J/$\psi$ and $\psi$' production in 450 GeV pA interactions and its dependence on the rapidity and $X_{F}$}, PhD Thesis, Instituto Superior Tecnico, Lisbon, Portugal (2001) {\tt[http://www.cern.ch/NA50/theses/ruben.ps.gz]}.
\bibitem{T_pPb} ALICE Collaboration, \emph{Transverse Momentum Distribution and Nuclear Modification Factor of Charged Particles in p-Pb Collisions at $\sqrt{s_{NN}}$ = 5.02 TeV}, \emph{Phys. Rev. Lett.} {\bf 110} (2013) 082302 {\tt[arXiv:1210.4520] [INSPIRE]}.
\bibitem{CS_interpolation} ALICE and LHCb Collaborations, \emph{Reference pp cross-sections for J/$\psi$ studies in proton-lead collisions at  $\sqrt{s_{NN}}$ = 5.02 TeV and comparisons between ALICE and LHCb results}, {\tt [ALICE-PUBLIC-2013-002] [LHCb-CONF-2013-013]}.
\bibitem{interpol_ele1} A. Adare, et al., \emph{J/$\psi$ Production vs Transverse Momentum and Rapidity in p+p Collisions at $\sqrt{s}$ = 200 GeV}, \emph{Phys.Rev.Lett.}, {\bf 98} (2007) 232002 {\tt [arXiv:hep-ex/0611020]}.
\bibitem{interpol_ele2} D. Acosta, et al., \emph{Measurement of the J/$\psi$ Meson and b-Hadron Production Cross Sections in ppbar Collisions at $\sqrt{s}$ = 1960 GeV}, \emph{Phys.Rev.D}, {\bf 71} (2005) 032001 {\tt [arXiv:hep-ex/0412071]}.
\bibitem{interpol_ele3} B. Abelev, et al., \emph{Inclusive J/$\psi$ production in pp collisions at $\sqrt{s}$ = 2.76 TeV}, \emph{Phys.Lett.B}, {\bf 718} (2012) 295–306 {\tt [arXiv:1203.3641]}.
\bibitem{interpol_ele4} K. Aamodt, et al., \emph{Rapidity and transverse momentum dependence of inclusive J/$\psi$ production in pp collisions at $\sqrt{s}$ = 7 TeV}, \emph{Phys.Lett.B}, {\bf 704} (2011) 442–455 {\tt [arXiv:1105.0380]}.
\bibitem{EPS09_1} J. Albacete et al., \emph{Predictions for p+Pb Collisions at $\sqrt{s_{NN}}$ = 5TeV}, \emph{Int. J. Mod. Phys.}, {\bf E 22} (2013) 1330007 {\tt [arXiv:1301.3395] [INSPIRE]}.
\bibitem{EPS09_2} R. Vogt, private communications.
\bibitem{ElossAlice} F. Arleo and S. Peigne, \emph{Heavy-quarkonium suppression in p-A collisions from parton energy loss in cold QCD matter}, \emph{JHEP}, {\bf 03} (2013) 122 {\tt [arXiv:1212.0434] [INSPIRE]}.
\bibitem{CGCAlice} H. Fujii and K. Watanabe, \emph{Heavy quark pair production in high energy pA collisions: Quarkonium}, \emph{Nucl. Phys.}, {\bf A 915} (2013) 1 {\tt [arXiv:1304.2221] [INSPIRE]}.
\bibitem{PHENIX} A. Adare et al. (PHENIX Collaboration), \emph{Nuclear Modification of $\psi$', $\chi_{c}$, and J/$\psi$ Production in d+Au Collisions at $\sqrt{s_{NN}}$ = 200 GeV}, \emph{Phys. Rev. Lett.}, {\bf 111} (2013) 202301.
\end{thebibliography}
\end{document}